\documentclass[english]{article}
\usepackage[T1]{fontenc}
\usepackage[latin9]{inputenc}
\usepackage{amsmath}
\usepackage{amsthm}
\usepackage{amssymb}
\usepackage{graphicx}

\makeatletter
\theoremstyle{plain}
\newtheorem{thm}{\protect\theoremname}
\theoremstyle{plain}
\newtheorem{lem}[thm]{\protect\lemmaname}

\usepackage{spconf}
\usepackage{cite}
\newtheorem{definition}{Definition}

\makeatother

\usepackage{babel}
\providecommand{\lemmaname}{Lemma}
\providecommand{\theoremname}{Theorem}

\begin{document}
\global\long\def\X#1{\mathbf{X}^{(#1)}}
\global\long\def\N#1{\mathbf{N}^{(#1)}}
\global\long\def\Xp#1{\mathbf{X}^{'(#1)}}
\global\long\def\x#1{x^{(#1)}}
\global\long\def\pt#1{\tilde{\mathbf{p}}_{#1,h}}
\global\long\def\gt{\tilde{\mathbf{G}}_{h}}
\global\long\def\g{\tilde{\mathbf{G}}}
\global\long\def\gtt#1{\tilde{\mathbf{G}}_{h,#1}}
\global\long\def\bias{\mathbb{B}}
\global\long\def\var{\mathbb{V}}
\global\long\def\bE{\mathbb{E}}
\global\long\def\ez#1{\mathbb{E}_{\mathbf{#1}}}
\global\long\def\et#1{\tilde{\mathbf{e}}_{#1,h}}
\global\long\def\gttl#1{\tilde{\mathbf{G}}_{h(l),#1}}

\twoauthors{Kevin R. Moon}
{Yale University \\
Department of Genetics \\
New Haven, Connecticut, U.S.A}
{Morteza Noshad, Salimeh Yasaei Sekeh, Alfred O. Hero III\sthanks{The research in this paper was partially supported by grant W911NF-15-1-0479 from the US Army Research Office.}}
{University of Michigan\\
Electrical Engineering and Computer Science\\
Ann Arbor, Michigan, U.S.A}

\title{Information Theoretic Structure Learning with Confidence}

\maketitle
\ninept
\begin{abstract}
Information theoretic measures (e.g. the Kullback Liebler divergence
and Shannon mutual information) have been used for exploring possibly
nonlinear multivariate dependencies in high dimension. If these dependencies
are assumed to follow a Markov factor graph model, this exploration
process is called structure discovery. For discrete-valued samples,
estimates of the information divergence over the parametric class
of multinomial models lead to structure discovery methods whose mean
squared error achieves parametric convergence rates as the sample
size grows. However, a naive application of this method to continuous
nonparametric multivariate models converges much more slowly. In this
paper we introduce a new method for nonparametric structure discovery
that uses weighted ensemble divergence estimators that achieve parametric
convergence rates and obey an asymptotic central limit theorem that
facilitates hypothesis testing and other types of statistical validation.
\end{abstract}

\section{Introduction}

Information theoretic measures such as mutual information (MI) can
be applied to measure the strength of multivariate dependencies between
random variables (RV). Such measures are used in many applications
including determining channel capacity~\cite{cover2012elements},
image registration~\cite{viola1997alignment}, independent subspace
analysis~\cite{pal2010estimation}, and independent component analysis~\cite{comon1994independent}.
MI has also been used for structure learning in graphical models (GM)~\cite{chow1968approximating},
which are factorizable multivariate distributions that are Markovian
according to a graph~\cite{lauritzen1996graphical}. GMs have been
used in fields such as bioinformatics, image processing, control theory,
social science, and marketing analysis. However, structure learning
for GMs remains an open challenge since the most general case requires
a combinatorial search over the space of all possible structures \cite{mohan2012structured,yuan2007model}
and nonparametric approaches have poor convergence rates as the number
of samples increases. This prevents reliable application of nonparametric
structure learning except for impractically large sample sizes. This
paper proposes a nonparametric MI-based ensemble estimator for structure
learning that achieves the parametric mean squared error (MSE) rate
when the densities are sufficiently smooth and admits a central limit
theorem (CLT) which enables us to perform hypothesis testing. We demonstrate
this estimator in multiple structure learning experiments.

Several structure learning algorithms have been proposed for parametric
GMs including discrete Markov random fields~\cite{kindermann1980markov},
Gaussian GMs~\cite{edwards2012introduction}, and Bayesian networks~\cite{pearl2014probabilistic}.
Recently, the authors of \cite{anandkumar2013learning} proposed learning
latent variable models from observed samples by estimating dependencies
between observed and hidden variables. Numerous other works have demonstrated
that latent tree models can be learned efficiently in high dimensions
(e.g. \cite{choi2011learning,mossel2007distorted}). 

We focus on two methods of nonparametric structure learning based
on ensemble MI estimation. The first method is the Chow-Liu (CL) algorithm
which constructs a first order tree from the MI of all pairs of RVs
to approximate the joint pdf~\cite{chow1968approximating}. Since
structure learning approaches can suffer from performance degradation
when the model does not match the true distribution, we propose hypothesis
testing via MI estimation to determine how well the tree structure
imposed by the CL algorithm approximates the joint distribution. The
second method learns the structure by performing hypothesis testing
on the MI of all pairs of RVs. An edge is assigned between two RVs
if the MI is statistically different from zero.

Accurate MI estimation is necessary for both methods. Estimating MI
is often difficult, especially in high dimensions when there is no
parametric model for the data. Nonparametric methods of estimating
MI have been proposed including $k$-nearest neighbor based methods~\cite{kraskov2004estimating,kozachenko1987sample}
and minimal spanning trees~\cite{khan2007relative}. However, the
MSE convergence rates of the latter estimator are currently unknown,
while the $k$-nn based methods achieve the parametric rate only when
the dimension of each of the RVs is less than 3~\cite{gao2016demystifying}.

Recent work has focused on the more general problem of nonparametric
divergence estimation including approaches based on optimal kernel
density estimators (KDE)~\cite{krishnamurthy2014divergence,singh2014exponential,kandasamy2015nonparametric}
and ensemble methods~\cite{sricharan2013ensemble,moon2014isit,moon2014nips,moon2016isit}.
While the optimal KDE-based approaches can achieve the parametric
MSE rate for smooth densities (i.e. the densities are at least $d$~\cite{singh2014exponential}
or $d/2$~\cite{krishnamurthy2014divergence,kandasamy2015nonparametric}
times differentiable where $d$ is the dimension of the data), they
can be difficult to construct near the density support boundary and
they require knowledge of the boundary. Also, for some types of divergence
functionals, these approaches require numerical integration which
requires many computations. 

In contrast, the ensemble estimators in~\cite{moon2014isit,moon2014nips,moon2016isit,sricharan2013ensemble}
vary the neighborhood size of nonparametric density estimators to
construct an ensemble of simple plug-in divergence or entropy estimators.
The final estimator is a weighted average of the ensemble of estimators
where the weights are chosen to decrease the bias with only a small
increase in the variance. Specifically, the ensemble estimator in~\cite{moon2016isit}
achieves the parametric MSE rate without any knowledge of the densities'
support set when the densities are $(d+1)/2$ times differentiable.

\section{Factor Graph Learning}

This paper focuses on learning a second-order product approximation
(i.e. a dependence tree) of the joint probability distribution of
the data. Let $\X i$ denote the $i$th component of a $d$-dimensional
random vector $\mathbf{X}$. We use similar notation to~\cite{chow1968approximating}
where the goal is to approximate the joint probability density $p(\mathbf{X})$
as 
\begin{equation}
p'\left(\mathbf{X}\right)=\prod_{i=1}^{d}p\left(\X{m_{i}}|\X{m_{j(i)}}\right),\label{eq:approx}
\end{equation}
where $0\leq j(i)<i,$ $\left(m_{1},\dots,m_{d}\right)$ is a (unknown)
permutation of $1,\,2,\,\dots d$, $p\left(\X i|\X 0\right)=p\left(\X i\right)$,
and $p\left(\X i|\X j\right)$ ($j\neq0$) is the conditional probability
density of $\X i$ given $\X j$.

The CL algorithm~\cite{chow1968approximating} provides an information
theoretic method for selecting the second-order terms in (\ref{eq:approx}).
It chooses the second-order terms that minimize the Kullback-Leibler
(KL) divergence between the joint density $p(\mathbf{X})$ and the
approximation $p'(\mathbf{X})$. This reduces to constructing the
maximal spanning tree where the edge weights correspond to the MI
between the RVs at the vertices~\cite{chow1968approximating}.

In practice, the pairwise MI between each pair of RVs is rarely known
and must be estimated from data. Thus accurate MI estimators are required.
Furthermore, while the sum of the pairwise MI gives a measure of the
quality of the approximation, it does not indicate if the approximation
is a sufficiently good fit or whether a different model should be
used. This problem can be framed as testing the hypothesis that $p'(\mathbf{X})=p(\mathbf{X})$
at a prescribed false positive level. This test can be performed using
MI estimation.

In addition, we propose that (\ref{eq:approx}) can be learned by
performing hypothesis testing on the MI of all pairs of RVs and assigning
an edge between two RVs if the MI is statistically different from
zero. To account for the multiple comparisons bias, we control the
false discovery rate (FDR)~\cite{zhu2005high}.

\section{Mutual Information Estimation}

\label{sec:MI}Information theoretic methods for learning nonlinear
structures require accurate estimation of MI and estimates of its
sample distribution for hypothesis testing. In this section, we extend
the ensemble divergence estimators given in~\cite{moon2016isit}
to obtain an accurate MI estimator and use the CLT to justify a large
sample Gaussian approximation to the sampling distribution. We consider
general MI functionals. Let $g:(0,\infty)\rightarrow\mathbb{R}$ be
a smooth functional, e.g. $g(u)=\ln u$ for Shannon MI or $g(u)=u^{\alpha}$,
with $\alpha\in[0,1]$, for R\'enyi MI. Then the pairwise MI between
$\X i$ and $\X j$ can be defined as 
\begin{equation}
G_{ij}=\int g\left(\frac{p\left(\x i\right)p\left(\x j\right)}{p\left(\x i,\x j\right)}\right)p\left(\x i,\x j\right)d\x id\x j.\label{eq:g1}
\end{equation}
For hypothesis testing, we are interested in the following 
\begin{equation}
G\left(p;p'\right)=\int g\left(\frac{p'(x)}{p(x)}\right)p(x)dx.\label{eq:g2}
\end{equation}

In this paper we focus only on the case where the RVs are continuous
with smooth densities. To extend the method of ensemble estimation
in~\cite{moon2016isit} to MI, we 1) define simple KDE-based plug-in
estimators, 2) derive expressions for the bias and variance of these
base estimators, and 3) then take a weighted average of an ensemble
of these simple plug-in estimators to decrease the bias based on the
expressions derived in step 2). To perform hypothesis testing on the
estimator of (\ref{eq:g2}), we use a CLT. Note that we cannot simply
extend the divergence estimation results in~\cite{moon2016isit}
to MI as \cite{moon2016isit} assumes that the random variables from
different densities are independent which may not be the case for
(\ref{eq:g1}) or (\ref{eq:g2}).

We first define the plug-in estimators. The conditional probability
density is defined as the ratio of the joint and marginal densities.
Thus the ratio within the $g$ functional in (\ref{eq:g2}) can be
represented as the ratio of the product of some joint densities with
two random variables and the product of marginal densities in addition
to $p$. For example, if $d=3$ and $p'(\mathbf{X})=p\left(\X 1|\X 2\right)p\left(\X 2|\X 3\right)p\left(\X 3\right)$,
then 
\begin{equation}
\frac{p'(\mathbf{X})}{p(\mathbf{X})}=\frac{p\left(\X 1,\X 2\right)p\left(\X 2,\X 3\right)}{p\left(\X 2\right)p\left(\X 1,\X 2,\X 3\right)}.\label{eq:example}
\end{equation}

For the KDEs, assume that we have $N$ i.i.d. samples $\left\{ \mathbf{X}_{1},\dots,\mathbf{X}_{N}\right\} $
available from the joint density $p\left(\mathbf{X}\right)$. The
KDE of $p(\mathbf{X}_{j})$ is 
\[
\pt X(\mathbf{X}_{j})=\frac{1}{Mh^{d}}\sum_{\substack{i=1\\
i\neq j
}
}K\left(\frac{\mathbf{X}_{j}-\mathbf{X}_{i}}{h}\right),
\]
where $K$ is a symmetric product kernel function, $h$ is the bandwidth,
and $M=N-1$. Define the KDEs $\pt{ik}\left(\X i_{j},\X k_{j}\right)$
and $\pt i\left(\X i_{j}\right)$ (for $p\left(\X i_{j},\X k_{j}\right)$
and $p\left(\X i_{j}\right)$, respectively) similarly. Let $\pt X^{'}(\mathbf{X}_{j})$
be defined using the KDEs for the marginal densities and the joint
densities with two random variables. For example, in the example given
in (\ref{eq:example}), we have 
\[
\pt X^{'}(\mathbf{X}_{j})=\frac{\pt{12}\left(\X 1_{j},\X 2_{j}\right)\pt{23}\left(\X 2_{j},\X 3_{j}\right)}{\pt 2\left(\X 2_{j}\right)}.
\]
For brevity, we use the same bandwidth and product kernel for each
of the KDEs although our method generalizes to differing bandwidths
and kernels. The plug-in MI estimator for (\ref{eq:g2}) is then
\[
\gt=\frac{1}{N}\sum_{j=1}^{N}g\left(\frac{\pt X^{'}(\mathbf{X}_{j})}{\pt X(\mathbf{X}_{j})}\right).
\]
The plug-in estimator $\gtt{ij}$ for (\ref{eq:g1}) is defined similarly.

To apply bias-reducing ensemble methods similar to~\cite{moon2016isit}
to the plug-in estimators $\gt$ and $\gtt{ij}$, we need to derive
their MSE convergence rates. As in~\cite{moon2016isit}, we assume
that 1) the density $p(\mathbf{X})$ and all other marginal densities
and pairwise joint densities are $s\geq2$ times differentiable and
the functional $g$ is infinitely differentiable; 2) $p(\mathbf{X})$
has bounded support set $\mathcal{S}$; 3) all densities are strictly
lower bounded on their support sets. Additionally, we make the same
assumption on the boundary of the support as in \cite{moon2016isit}:
4) the support is smooth wrt the kernel $K(u)$ in the sense that
the expectation of the area outside of $\mathcal{S}$ wrt any RV $u$
with smooth distribution is a smooth function of the bandwidth $h$.
This assumption is satisfied, for example, when $\mathcal{S}$ is
the unit cube and $K(x)$ is the uniform rectangular kernel. For full
technical details on the assumptions, see Appendix~\ref{sec:assumptions}. 
\begin{thm}
\label{thm:bias}If $g$ is infinitely differentiable, then the bias
of $\gtt{ij}$ and $\gt$ are 
\begin{align}
\bias\left[\gtt{ij}\right] & =\sum_{m=1}^{\left\lfloor s\right\rfloor }c_{5,i,j,m}h^{m}+O\left(\frac{1}{Nh^{2}}+h^{s}\right)\nonumber \\
\bias\left[\gt\right] & =\sum_{m=1}^{\left\lfloor s\right\rfloor }c_{6,m}h^{m}+O\left(\frac{1}{Nh^{d}}+h^{s}\right).\label{eq:bias1}
\end{align}
If $g\left(t_{1}/t_{2}\right)$ has $k$, $l$-th order mixed derivatives
$\frac{\partial^{k+l}g(t_{1}/t_{2})}{\partial t_{1}^{k}\partial t_{2}^{l}}$
that depend on $t_{1}$, $t_{2}$ only through $t_{1}^{\alpha}t_{2}^{\beta}$
for some $\alpha,\,\beta\in\mathbb{R}$ for each $1\leq k,l\leq\lambda$
then the bias of $\gt$ is 
\begin{align}
\bias\left[\gt\right] & =\sum_{m=1}^{\left\lfloor s\right\rfloor }c_{6,m}h^{m}+\sum_{m=0}^{\left\lfloor s\right\rfloor }\sum_{q=1}^{\left\lfloor \lambda/2\right\rfloor }\left(\frac{c_{7,1,q,m}}{\left(Nh^{d}\right)^{q}}+\frac{c_{7,2,q,m}}{\left(Nh^{2}\right)^{q}}\right)h^{m}\nonumber \\
 & +O\left(\frac{1}{\left(Nh^{d}\right)^{\lambda/2}}+h^{s}\right).\label{eq:bias2}
\end{align}
\end{thm}
The expression in (\ref{eq:bias2}) allows us to achieve the parametric
MSE rate under less restrictive assumptions on the smoothness of the
densities ($s>d/2$ for (\ref{eq:bias2}) compared to $s\geq d$ for
(\ref{eq:bias1})). The extra condition required on the mixed derivatives
of $g$ to obtain the expression in (\ref{eq:bias2}) is satisfied,
for example, for Shannon and R\'enyi information measures. Note that
a similar expression could be derived for the bias of $\gtt{ij}$.
However, since $s\geq2$ is required and the largest dimension of
the densities estimated in $\gtt{ij}$ is 2, we would not achieve
any theoretical improvement in the convergence rate.
\begin{thm}
\label{thm:variance}If the functional $g(t_{1}/t_{2})$ is Lipschitz
continuous in both of its arguments with Lipschitz constant $C_{g}$,
then the variance of both $\gt$ and $\gtt{ij}$ is $O(1/N)$.
\end{thm}
The Lipschitz assumption on $g$ is comparable to assumptions required
by other nonparametric distributional functional estimators~\cite{moon2016isit,krishnamurthy2014divergence,kandasamy2015nonparametric,singh2014exponential}
and is ensured for functionals such as Shannon and R\'enyi informations
by our assumption that the densities are bounded away from zero. The
proofs of Theorems \ref{thm:bias} and \ref{thm:variance} share some
similarities with the bias and variance proofs for the divergence
functional estimators in \cite{moon2016isit}. The primary differences
deal with the product of KDEs. See the appendices\textbf{ }for the
full proofs.

From Theorems~\ref{thm:bias} and \ref{thm:variance}, letting $h\rightarrow0$
and $Nh^{2}\rightarrow\infty$ or $Nh^{d}\rightarrow\infty$ is required
for the respective MSE of $\gtt{ij}$ and $\gt$ to go to zero. Without
bias correction, the optimal MSE rate is, respectively, $O\left(N^{-2/3}\right)$
and $O\left(N^{-2/(d+1)}\right)$. Using an optimally weighted ensemble
of estimators enables us to perform bias correction and achieve much
better (parametric) convergence rates~\cite{moon2016isit,sricharan2013ensemble}.

The ensemble of estimators is created by varying the bandwidth $h$.
Choose $\bar{l}=\left\{ l_{1},\dots,l_{L}\right\} $ to be a set of
positive real numbers and let $h(l)$ be a function of the parameter
$l\in\bar{l}$. Define $w=\left\{ w(l_{1}),\dots,w(l_{L})\right\} $
and $\g_{w}=\sum_{l\in\bar{l}}w(l)\g_{h(l)}$. Theorem 4 in \cite{moon2016isit}
indicates that if enough of the terms in the bias expression of an
estimator within an ensemble of estimators are known and the variance
is $O(1/N)$, then the weight $w_{0}$ can be chosen so that the MSE
rate of $\g_{w_{0}}$ is $O(1/N)$, i.e. the parametric rate. The
theorem can be applied as follows. For general $g$, let $h(l)=lN^{-1/(2d)}$
for $\g_{h(l)}$. Denote $\psi_{m}(l)=l^{m}$ with $m\in J=\{1,\dots,\left\lfloor s\right\rfloor \}$.
The optimal weight $w_{0}$ is obtained by solving 
\begin{equation}
\begin{array}{rl}
\min_{w} & ||w||_{2}\\
subject\,to & \sum_{l\in\bar{l}}w(l)=1,\\
 & \left|\sum_{l\in\bar{l}}w(l)\psi_{m}(l)\right|=0,\,\,m\in J,
\end{array}\label{eq:opt}
\end{equation}
It can then be shown that the MSE of $\g_{w_{0}}$ is $O(1/N)$ as
long as $s\geq d$. This works by using the last line in (\ref{eq:opt})
to cancel the lower-order terms in the bias. Similarly, by using the
same optimization problem we can define a weighted ensemble estimator
$\g_{w_{0},ij}$ of $G_{ij}$ that achieves the parametric rate when
$h(l)=lN^{-1/4}$ which results in $\psi_{m}(l)=l^{m}$ for $m\in J=\{1,2\}$.
These estimators correspond to the ODin1 estimators defined in~\cite{moon2016isit}.

An ODin2 estimator of $G\left(p;p'\right)$ can be derived using (\ref{eq:bias2}).
Let $\delta>0$, assume that $s\geq(d+\delta)/2$, and let $h(l)=lN^{-1/(d+\delta)}$.
This results in the function $\psi_{1,m,q}(l)=l^{m-dq}$ for $m\in\left\{ 0,\dots,(d+\delta)/2\right\} $
and $q\in\{0,\dots,(d+\delta)/\delta\}$ with the restriction that
$m+q\neq0$. Additionally we have $\psi_{2,m,q}(l)=l^{m-2q}$ for
$m\in\{0,\dots,(d+\delta)/2\}$ and $q\in\{1,\dots,(d+\delta)/(2(d+\delta-2))\}$.
These functions correspond to the lower order terms in the bias. Then
using (\ref{eq:opt}) with these functions results in a weight vector
$w_{0}$ such that $\g_{w_{0}}$ achieves the parametric rate as long
as $s\geq(d+\delta)/2$. Then since $\delta$ is arbitrary, we can
achieve the parametric rate for $s>d/2$. 

We conclude this section by giving a CLT. This theorem provides justification
for performing structural hypothesis testing with the estimators $\g_{w_{0}}$
and $\g_{w_{0},ij}$. The proof uses an application of Slutsky's Theorem
preceded by the Efron-Stein inequality that is similar to the proof
of the CLT of the divergence ensemble estimators in~\cite{moon2016isit}.
The extension of the CLT in \cite{moon2016isit} to $\g_{w}$ is analogous
to the extension required in the proof of the variance results in
Theorem~\ref{thm:variance}.
\begin{thm}
\label{thm:CLT}Assume that $h=o(1)$ and $Nh^{d}\rightarrow\infty$.
If $\mathbf{S}$ is a standard normal random variable, $L$ is fixed,
and $g$ is Lipschitz in both arguments, then 
\[
\Pr\left(\left(\g_{w}-\bE\left[\g_{w}\right]\right)/\sqrt{\var\left[\g_{w}\right]}\leq t\right)\rightarrow\Pr(\mathbf{S}\leq t).
\]
\end{thm}

\section{Experiments}

We perform multiple experiments to demonstrate the utility of our
proposed methods for structure learning of a GM with $d=3$ nodes
whose structure is a nonlinear Markov chain from nodes $i=1$ to $i=2$
to $i=3$. That is, out of a possible 6 edges in a complete graph,
only the node pairs $(1,2)$ and $(2,3)$ are connected by edges.\textbf{
}In all experiments, $\X 1\sim\text{Unif}(-0.5,0.5)$, $\N i\sim\mathcal{N}(0,0.5)$,
and $\N 1$ and $\N 2$ are independent. We have $N=500$ i.i.d. samples
from $\X 1$ and choose an ensemble of bandwidth parameters with $L=50$
based on the guidelines in~\cite{moon2016isit}. To better control
the variance, we calculate the weight $w_{0}$ using the relaxed version
of (\ref{eq:opt}) given in~\cite{moon2016isit}. We compare the
results of the ensemble estimators ODin1 and ODin2 ($\delta=1$ in
the latter) to the simple plug-in KDE estimator. All $p$-values are
constructed by applying Theorem~\ref{thm:CLT} where the mean and
variance of the estimators are estimated via bootstrapping. In addition,
we studentize the data at each node by dividing by the sample standard
deviation as is commonly done in entropic machine learning. This introduces
some dependency between the nodes that decreases as $O\left(1/N\right)$.
This studentization has the effect of reducing the dependence of the
MI on the marginal distributions and stabilizing the MI estimates.
We estimate the R\'enyi-$\alpha$ integral for R\'enyi MI with $\alpha=0.5$;
i.e. $g(u)=u^{\alpha}$. Thus if the ratio of densities within (\ref{eq:g1})
or (\ref{eq:g2}) is 1, the R\'enyi-$\alpha$ integral is also 1. 

In the first type of experiments, we vary the signal-to-noise ratio
(SNR) of a Markov chain by varying the parameter $a$ and setting
\begin{align}
\X 2 & =\left(\X 1\right)^{2}+a\N 1,\nonumber \\
\X 3 & =\left(\X 2\right)^{2}+a\N 2.\label{eq:exp1}
\end{align}
In the second type of experiments, we create a cycle within the graph
by setting
\begin{align}
\X 2 & =\left(\X 1\right)^{2}+a\N 1,\nonumber \\
\X 3 & =\left(\X 2\right)^{2}+b\X 1+a\N 2.\label{eq:exp2}
\end{align}
We either fix $b$ and vary $a$ or vice versa.

\begin{figure}
\centering

\includegraphics[width=0.5\textwidth]{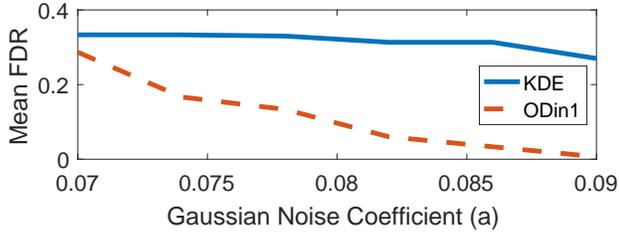}

\caption{\label{fig:fdr}The mean FDR from 100 trials as a function of $a$
when estimating the MI between all pairs of RVs for (\ref{eq:exp1})
with significance level $\gamma=0.1$. The dependence between $\protect\X 1$
and $\protect\X 3$ decreases as the noise increases resulting in
lower mean FDR.}
\end{figure}

We first use hypothesis testing on the estimated pairwise MI to learn
the structure in (\ref{eq:exp1}). We do this by testing the null
hypotheses that each pairwise R\'enyi-$\alpha$ integral is equal
to 1. We do not use the ODin2 estimator in this experiment as there
is no theoretical gain in MSE over ODin1 for pairwise MI estimation.
Figure~\ref{fig:fdr} plots the mean FDR from 100 trials as a function
of $a$ under this setting with significance level $\gamma=0.1$.
In ths case, the FDR is either $0$ (no false discoveries) or $1/3$
(one false discovery). Thus the mean FDR provides an indicator for
the number of trials where a false discovery occurs. Figure~\ref{fig:fdr}
shows that the mean FDR decreases slowly for the KDE estimator and
rapidly for the ODin1 estimator as the noise increases. Since $\X 3$
is a function of $\X 2$ which is a function of $\X 1$, then $G_{13}\neq1$.
However, as the noise increases, the relative dependence of $\X 3$
on $\X 1$ decreases and thus $G_{13}$ approaches 1. The ODin1 estimator
tracks this approach better as the corresponding FDR decreases at
a faster rate compared to the KDE estimator.

\begin{figure}
\centering

\includegraphics[width=0.5\textwidth]{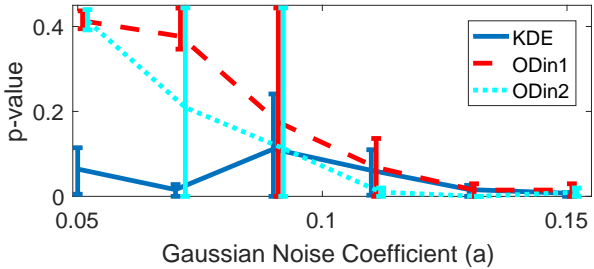}

\caption{\label{fig:fdrCL}The average $p$-value with error bars at the 20th
and 80th percentiles from 90 trials for the hypothesis test that $G(p;p')=1$
after running the CL algorithm for (\ref{eq:exp1}). The graphs are
offset horizontally for better visualization. Higher noise levels
lead to higher error rates in the CL algorithm and thus lower $p$-values.}
\end{figure}

In the next set of experiments, the CL algorithm estimates the tree
structure in (\ref{eq:exp1}) and we test the hypothesis that $G(p;p')=1$
to determine if the output of the CL algorithm gives the correct structure.
The resulting mean $p$-value with error bars at the 20th and 80th
percentiles from 90 trials is given in Figure~\ref{fig:fdrCL}. High
$p$-values indicate that both the CL algorithm performs well and
that $G(p;p')$ is not statistically different from 1. The ODin1 estimator
generally has higher values than the ODin2 and KDE estimators which
indicates better performance.

\begin{figure}
\centering

\includegraphics[width=0.5\textwidth]{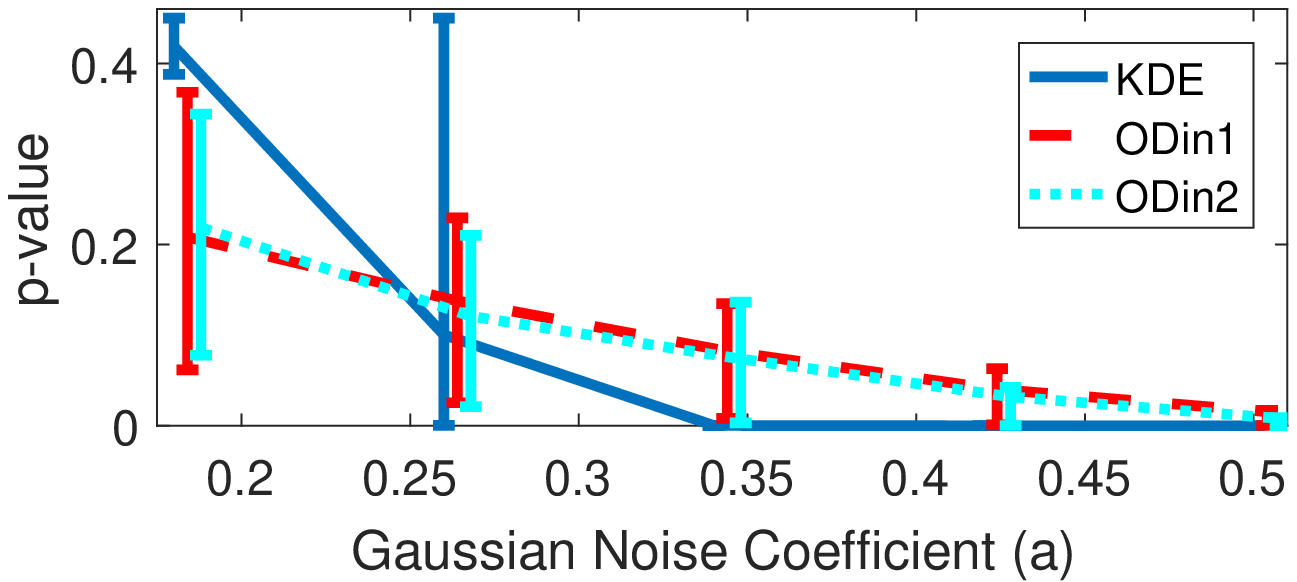}

\includegraphics[width=0.5\textwidth]{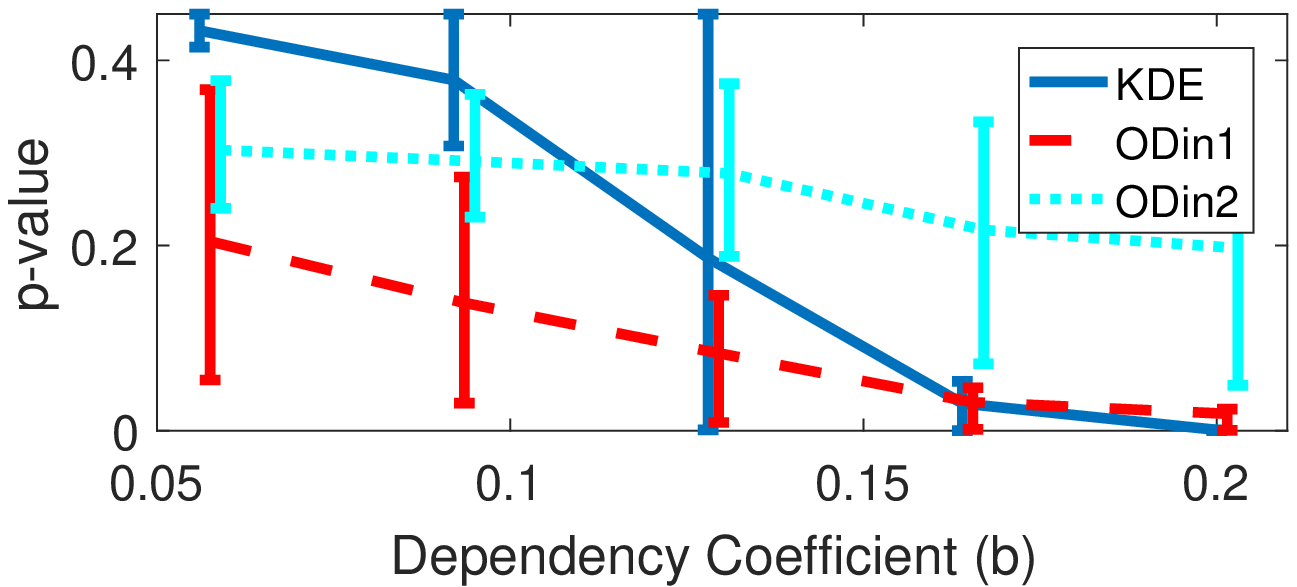}

\caption{\label{fig:incorrectCL}The mean $p$-value with error bars at the
20th and 80th percentiles from 100 trials for the hypothesis test
that $G\left(p;p'\right)=1$ for (\ref{eq:exp2}) when the CL tree
does not give the correct structure. Top: $b=0.5$ and $a$ varies.
Bottom: $a=0.05$ and $b$ varies. The graphs are offset horizontally
for better visualization. Low $p$-values indicate better performance.
The ODin1 estimator generally matches or outperforms the other estimators,
especially in the lower noise cases.}
\end{figure}

The final set of experiments focuses on (\ref{eq:exp2}). In this
case, the CL tree does not include the edge between $\X 1$ and $\X 3$
and we report the $p$-values for the hypothesis that $G\left(p;p'\right)=1$
when varying either $a$ or $b$. The mean $p$-value with error bars
at the 20th and 80th percentiles from 100 trials are given in Figure~\ref{fig:incorrectCL}.
In the top figure, we fix $b=0.5$ and vary the noise parameter $a$
while in the bottom figure we fix $a=0.05$ and vary $b$. Thus the
true structure does not match the CL tree and low $p$-values are
desired. For low noise in the top figure (fixed dependency coefficient),
the ODin estimators perform better than the KDE estimator and have
less variability. In the bottom figure (fixed noise), the ODin1 estimator
generally outperforms the other estimators.

In general, the ODin1 estimator outperforms the other estimators in
these experiments. Future work includes investigating other scenarios
including larger dimensional data and larger sample sizes. Based on
the experiments in~\cite{moon2016arxiv,moon2016isit}, it is possible
that the ODin2 estimator will perform comparably to the ODin1 estimator
and that both ODin estimators will outperform the KDE estimator in
higher dimensions.

\section{Conclusion}

We derived the convergence rates for a kernel density plug-in estimator
of MI functionals and proposed nonparametric ensemble estimators with
a CLT that achieve the parametric rate when the densities are sufficiently
smooth. We proposed two approaches for hypothesis testing based on
the CLT to learn the structure of the data. The experiments demonstrated
the utility of these approaches in structure learning and demonstrated
the improvement of ensemble methods over the plug-in method for a
low dimensional example. 

\bibliographystyle{IEEEbib}
\bibliography{References}

\appendix
\onecolumn

\section{Assumptions}

\label{sec:assumptions}In the proofs, we assume the more general
H\"{o}lder condition of smoothness on the densities: 

\begin{definition}[H\"{o}lder Class] \label{def:holder}\emph{Let
$\mathcal{X}\subset\mathbb{R}^{d}$ be a compact space. For $r=(r_{1},\dots,r_{d}),$
$r_{i}\in\mathbb{N},$ define $|r|=\sum_{i=1}^{d}r_{i}$ and $D^{r}=\frac{\partial^{|r|}}{\partial x_{1}^{r_{1}}\dots\partial x_{d}^{r_{d}}}$.
The H\"{o}lder class $\Sigma(s,H)$ of functions on $L_{2}(\mathcal{X})$
consists of the functions $f$ that satisfy 
\[
\left|D^{r}f(x)-D^{r}f(y)\right|\leq H\left\Vert x-y\right\Vert ^{s-r},
\]
for all $x,\,y\in\mathcal{X}$ and for all $r$ s.t. $|r|\leq\left\lfloor s\right\rfloor $.
}\end{definition}

Note that if $p\in\Sigma(s,H)$, then $p$ has at least $\left\lfloor s\right\rfloor $
derivatives.

The full assumptions we make on the densities and the functional $g$,
which we adapt from~\cite{moon2016arxiv}, are 
\begin{itemize}
\item $(\mathcal{A}.0)$: Assume that the kernel $K$ is a symmetric product
kernel with bounded support in each dimension.
\item $(\mathcal{A}.1)$: Assume there exist constants $\epsilon_{0},\epsilon_{\infty}$
such that $0<\epsilon_{0}\leq p(x)\leq\epsilon_{\infty}<\infty,\,\forall x\in\mathcal{S}$
and similarly that the marginal densities and joint pairwise densities
are bounded above and below.
\item $(\mathcal{A}.2)$: Assume that each of the densities belong to $\Sigma(s,H)$
in the interior of their support sets with $s\geq2$.
\item $(\mathcal{A}.3)$: Assume that $g\left(t_{1}/t_{2}\right)$ has an
infinite number of mixed derivatives wrt $t_{1}$ and $t_{2}$.
\item $(\mathcal{A}.4$): Assume that $\left|\frac{\partial^{k+l}g(t_{1},t_{2})}{\partial t_{1}^{k}\partial t_{2}^{l}}\right|/(k!l!)$,
$k,l=0,1,\ldots$ are strictly upper bounded for $\epsilon_{0}\leq t_{1},t_{2}\leq\epsilon_{\infty}$. 
\item $(\mathcal{A}.5)$: Assume the following boundary smoothness condition:
Let $p_{x}(u):\mathbb{R}^{d}\rightarrow\mathbb{R}$ be a polynomial
in $u$ of order $q\leq r=\left\lfloor s\right\rfloor $ whose coefficients
are a function of $x$ and are $r-q$ times differentiable. Then assume
that 
\[
\int_{x\in\mathcal{S}}\left(\int_{u:K(u)>0,\,x+uh\notin\mathcal{S}}K(u)p_{x}(u)du\right)^{t}dx=v_{t}(h),
\]
where $v_{t}(h)$ admits the expansion 
\[
v_{t}(h)=\sum_{i=1}^{r-q}e_{i,q,t}h^{i}+o\left(h^{r-q}\right),
\]
for some constants $e_{i,q,t}$.
\end{itemize}
It has been shown that assumption $\mathcal{A}.5$ is satisfied when
$\mathcal{S}$ is rectangular (e.g. $\mathcal{S}=[-1,1]^{d}$) and
$K$ is the uniform rectangular kernel~\cite{moon2016arxiv}. Thus
it can be applied to any densities in $\Sigma(s,H)$ on this support.

\section{Proof of Bias Results}

We prove Theorem \ref{thm:bias} in this appendix. The proof shares
some similarities with the bias proof of the divergence functional
estimators in~\cite{moon2016arxiv}. The primary differences lie
in handling the possible dependencies between random variables. We
focus on the more difficult case of $\gt$ as the bias derivation
for $\gtt{ij}$ is similar. 

Recall that $\pt X^{'}$ is a ratio of two products of KDEs. The numerator
is a product of 2-dimensional KDEs while the denominator is a product
of marginal (1-dimensional) KDEs, all with the same bandwidth. Let
$\gamma\subset\left\{ (i,j):i,j\in\left\{ 1,\dots,d\right\} \right\} $
denote the set of index pairs that denote the components of $\mathbf{X}$
that have joint KDEs that are in the product in the numerator of $\pt X^{'}$.
Let $\beta$ denote the set of indices that denote the components
of $\mathbf{X}$ that have marginal KDEs that are in the product in
the denominator of $\pt X^{'}$. Note that $|\gamma|=d-1$ and $|\beta|=d-2$.
As an example, in the example given in (\ref{eq:example}), we have
$\gamma=\left\{ (1,2),(2,3)\right\} $ and $\beta=\{2\}$. The bias
of $\gt$ can then be expressed as 
\begin{align}
\bias\left[\gt\right] & =\bE\left[g\left(\frac{\pt X^{'}(\mathbf{X})}{\pt X(\mathbf{X})}\right)-g\left(\frac{p'(\mathbf{X})}{p(\mathbf{X})}\right)\right]\nonumber \\
 & =\bE\left[g\left(\frac{\pt X^{'}(\mathbf{X})}{\pt X(\mathbf{X})}\right)-g\left(\frac{\prod_{(i,j)\in\gamma}\ez X\left[\pt{ij}\left(\X i,\X j\right)\right]}{\ez X\left[\pt X(\mathbf{X})\right]\prod_{k\in\beta}\ez X\left[\pt k\left(\X k\right)\right]}\right)\right]\nonumber \\
 & +\bE\left[g\left(\frac{\prod_{(i,j)\in\gamma}\ez X\left[\pt{ij}\left(\X i,\X j\right)\right]}{\ez X\left[\pt X(\mathbf{X})\right]\prod_{k\in\beta}\ez X\left[\pt k\left(\X k\right)\right]}\right)-g\left(\frac{p'(\mathbf{X})}{p(\mathbf{X})}\right)\right],\label{eq:gsplit}
\end{align}
where $\mathbf{X}$ is drawn from $p$ and $\ez X$ denotes the conditional
expectation given $\mathbf{X}$. We can view these terms as a variance-like
component (the first term) and a bias-like component, where the respective
Taylor series expansions depend on variance-like or bias-like terms
of the KDEs.

We first consider the bias-like term, i.e. the second term in (\ref{eq:gsplit}).
The Taylor series expansion of $g\left(\frac{\prod_{(i,j)\in\gamma}\ez X\left[\pt{ij}\left(\X i,\X j\right)\right]}{\ez X\left[\pt X(\mathbf{X})\right]\prod_{k\in\beta}\ez X\left[\pt k\left(\X k\right)\right]}\right)$
around $\prod_{(i,j)\in\gamma}p\left(\X i,\X j\right)$ and $p(\mathbf{X})\prod_{k\in\gamma}p(\X k)$
gives an expansion with terms of the form of 
\begin{align*}
\bias_{\mathbf{X}}^{m}\left[\prod_{(i,j)\in\gamma}\pt{ij}\left(\X i,\X j\right)\right] & =\left(\prod_{(i,j)\in\gamma}\ez X\left[\pt{ij}\left(\X i,\X j\right)\right]-\prod_{(i,j)\in\gamma}p\left(\X i,\X j\right)\right)^{m},\\
\bias_{\mathbf{X}}^{m}\left[p(\mathbf{X})\prod_{k\in\gamma}p(\X k)\right] & =\left(\ez X\left[\pt X(\mathbf{X})\right]\prod_{k\in\beta}\ez X\left[\pt k\left(\X k\right)\right]-p(\mathbf{X})\prod_{k\in\gamma}p(\X k)\right)^{m}.
\end{align*}

Since we are not doing boundary correction, we need to consider separately
the cases when $\mathbf{X}$ is in the interior of the support $\mathcal{S}$
and when $\mathbf{X}$ is close to the boundary of the support. For
precise definitions, a point $X\in\mathcal{S}$ is in the interior
of $\mathcal{S}$ if for all $X^{'}\notin\mathcal{S}$, $K\left(\frac{X-X^{'}}{h}\right)=0$,
and a point $X\in\mathcal{S}$ is near the boundary of the support
if it is not in the interior. Since $K$ is a product kernel, $X\in\mathcal{S}$
is in the interior if and only if all of the components of $X$ are
in their respective interiors. 

Assume that $\mathbf{X}$ is drawn from the interior of $\mathcal{S}$.
By a Taylor series expansion of the probability density $p$, we have
that 
\begin{align}
\ez X\left[\pt X(\mathbf{X})\right] & =\frac{1}{h^{d}}\int K\left(\frac{\mathbf{X}-x}{h}\right)p\left(x\right)dx\nonumber \\
 & =\int K(u)p(\mathbf{X}-uh)du\nonumber \\
 & =p(\mathbf{X})+\sum_{j=1}^{\left\lfloor s/2\right\rfloor }c_{X,j}(\mathbf{X})h^{2j}+O\left(h^{s}\right).\label{eq:E_p}
\end{align}
Similar expressions can be obtained for $\ez X\left[\pt{ij}\left(\X i,\X j\right)\right]$
and $\ez X\left[\pt k\left(\X k\right)\right]$.

Now assume that $\mathbf{X}$ lies near the boundary of the support
$\mathcal{S}$. In this case, we extend the expectation beyond the
support of the density: 
\begin{align}
\ez X\left[\pt X(\mathbf{X})\right]-p(\mathbf{X}) & =\frac{1}{h^{d}}\int_{x:x\in\mathcal{S}}K\left(\frac{\mathbf{X}-x}{h}\right)p(x)dx-p(\mathbf{X})\nonumber \\
 & =\left[\frac{1}{h^{d}}\int_{x:K\left(\frac{\mathbf{X}-x}{h}\right)>0}K\left(\frac{\mathbf{X}-x}{h}\right)p(x)dx-p(\mathbf{X})\right]\nonumber \\
 & -\left[\frac{1}{h^{d}}\int_{x:x\notin\mathcal{S}}K\left(\frac{\mathbf{X}-x}{h}\right)p(x)dx\right]\nonumber \\
 & =T_{1,X}(\mathbf{X})-T_{2,X}(\mathbf{X}).\label{eq:Tdiff}
\end{align}
We only evaluate the density $p$ and its derivatives at points within
the support when we take its Taylor series expansion. Thus the exact
manner in which we define the extension of $p$ does not matter as
long as the Taylor series remains the same and as long as the extension
is smooth. Thus the expected value of $T_{1,X}(\mathbf{X})$ gives
an expression of the form of (\ref{eq:E_p}). For the $T_{2,X}(\mathbf{X})$
term, we perform a similar Taylor series expansion and then apply
the condition in assumption $\mathcal{A}.5$ to obtain
\[
\bE\left[T_{2,X}(\mathbf{X})\right]=\sum_{i=1}^{r}e_{i}h^{i}+o\left(h^{r}\right).
\]
Similar expressions can be found for $\pt{ij}\left(\X i,\X j\right)$
, $\pt k\left(\X k\right)$, and when (\ref{eq:Tdiff}) is raised
to a power $t$. Applying this result gives for the second term in
(\ref{eq:gsplit}), 
\begin{equation}
\sum_{j=1}^{r}c_{g,p,j}h^{j}+O\left(h^{s}\right),\label{eq:h_boundary}
\end{equation}
where the constants $c_{g,p,j}$ depend on the densities, their derivatives,
and the functional $g$ and its derivatives.

For the first term in (\ref{eq:gsplit}), a Taylor series expansion
of $g\left(\frac{\pt X^{'}(\mathbf{X})}{\pt X(\mathbf{X})}\right)$
around $\prod_{(i,j)\in\gamma}\ez X\left[\pt{ij}\left(\X i,\X j\right)\right]$
and\\ $\ez X\left[\pt X(\mathbf{X})\right]\prod_{k\in\beta}\ez X\left[\pt k\left(\X k\right)\right]$
gives an expansion with terms of the form of 
\begin{eqnarray*}
\et 1^{q}(\mathbf{X}) & = & \left(\prod_{(i,j)\in\gamma}\pt{ij}\left(\X i,\X j\right)-\prod_{(i,j)\in\gamma}\ez X\left[\pt{ij}\left(\X i,\X j\right)\right]\right)^{q},\\
\et 2^{q}(\mathbf{X}) & = & \left(\pt X(\mathbf{X})\prod_{k\in\beta}\pt k\left(\X k\right)-\ez X\left[\pt X(\mathbf{X})\right]\prod_{k\in\beta}\ez X\left[\pt k\left(\X k\right)\right]\right)^{q}.
\end{eqnarray*}
To control these terms, we need expressions for $\ez X\left[\et 1^{q}(\mathbf{X})\right]$
and $\ez X\left[\et 2^{q}(\mathbf{X})\right]$. We'll derive the expression
only for $\ez X\left[\et 1^{q}(\mathbf{X})\right]$ as the expression
for $\ez X\left[\et 2^{q}(\mathbf{X})\right]$ is obtained in a similar
manner. 

For simplicity of exposition, we assume that $d=3$ and that $\gamma=\left\{ (1,2),(2,3)\right\} $.
Note that our method extends easily to the general case where notation
can be cumbersome. Define 
\begin{align*}
\mathbf{V}_{i,j}(\mathbf{X}) & =K_{1}\left(\frac{\X 1_{i}-\X 1}{h}\right)K_{2}\left(\frac{\X 2_{i}-\X 2}{h}\right)K_{2}\left(\frac{\X 2_{j}-\X 2}{h}\right)K_{3}\left(\frac{\X 3_{j}-\X 3}{h}\right)\\
 & -\ez X\left[K_{1}\left(\frac{\X 1_{i}-\X 1}{h}\right)K_{2}\left(\frac{\X 2_{i}-\X 2}{h}\right)\right]\ez X\left[K_{2}\left(\frac{\X 2_{j}-\X 2}{h}\right)K_{3}\left(\frac{\X 3_{j}-\X 3}{h}\right)\right]\\
 & =\mathbf{\eta}_{ij}(\mathbf{X})-\ez X\left[\mathbf{\eta}_{i}(\mathbf{X})\right]\ez X\left[\mathbf{\eta}_{j}^{'}(\mathbf{X})\right].
\end{align*}
Therefore, 
\[
\et 1(\mathbf{X})=\frac{1}{\left(Nh^{2}\right)^{|\gamma|}}\sum_{i=1}^{N}\sum_{j=1}^{N}\mathbf{V}_{i,j}(\mathbf{X}).
\]
By the binomial theorem, 
\[
\ez X\left[\mathbf{V}_{i,j}^{k}(\mathbf{X})\right]=\sum_{l=0}^{k}\binom{l}{k}\ez X\left[\mathbf{\eta}_{ij}^{l}(\mathbf{X})\right]\left[\ez X\left[\eta_{i}(\mathbf{X})\right]\ez X\left[\eta_{j}^{'}(\mathbf{X})\right]\right]^{k-l}.
\]
 It can then be seen using a similar Taylor Series analysis as in
the derivation of (\ref{eq:E_p}) that for $\mathbf{X}$ in the interior
of $\mathcal{S}$ and $i\neq j$, we have that 
\[
\ez X\left[\mathbf{\eta}_{ij}^{l}(\mathbf{X})\right]=h^{2|\gamma|}\sum_{m=0}^{\left\lfloor s/2\right\rfloor }c_{2,1,m,l}(\mathbf{X})h^{2m}.
\]
Combining these results gives for $i\neq j$
\[
\ez X\left[\mathbf{V}_{i,j}^{k}(\mathbf{X})\right]=h^{2|\gamma|}\sum_{m=0}^{\left\lfloor s/2\right\rfloor }c_{2,2,m,k}(\mathbf{X})h^{2m}+O\left(h^{4|\gamma|}\right).
\]
If $i=j$, we obtain 
\[
\ez X\left[\mathbf{\eta}_{ii}^{l}(\mathbf{X})\right]=h^{d}\sum_{m=0}^{\left\lfloor s/2\right\rfloor }c_{2,2,m}(\mathbf{X})h^{2m}.
\]
This then gives 
\[
\ez X\left[\mathbf{V}_{i,i}^{k}(\mathbf{X})\right]=h^{d}\sum_{m=0}^{\left\lfloor s/2\right\rfloor }c_{2,m,k}(\mathbf{X})h^{2m}+O\left(h^{4|\gamma|}\right)
\]
 Here the constants $c_{2,i,m,k}(\mathbf{X})$ depend on the density
$p$, its derivatives, and the moments of the kernels. 

Let $n(q)$ be the set of integer divisors of $q$ including 1 but
excluding $q$. Then due to the independence of the different samples,
it can then be shown that 
\begin{equation}
\ez X\left[\et 1^{q}(\mathbf{X})\right]=\sum_{i\in n(q)}\sum_{m=0}^{\left\lfloor s/2\right\rfloor }\left(\frac{c_{3,1,m,q}(\mathbf{X})}{\left(Nh^{2}\right)^{(q-i)}}+\frac{c_{3,2,m,q}(\mathbf{X})}{\left(Nh\right)^{(q-i)}}\right)h^{2m}+o\left(\frac{1}{N}\right).\label{eq:varterms1}
\end{equation}
By a similar procedure, we can show that 
\begin{equation}
\ez X\left[\et 2^{q}(\mathbf{X})\right]=\sum_{i\in n(q)}\sum_{m=0}^{\left\lfloor s/2\right\rfloor }\left(\sum_{j=0}^{|\beta|}\frac{c_{4,1,j,m,q}(\mathbf{X})}{\left(Nh^{d}\left(Nh\right)^{j}\right)^{q-i}}+\frac{c_{4,2,m,q}(\mathbf{X})}{\left(Nh\right)^{q-i}}\right)h^{2m}+o\left(\frac{1}{N}\right).\label{eq:varterms2}
\end{equation}
When $\mathbf{X}$ is near the boundary of the support, we can obtain
similar expressions as in (\ref{eq:varterms1}) and (\ref{eq:varterms2})
by following a similar procedure as in the derivation of (\ref{eq:h_boundary}).
The primary difference is that we will then have powers of $h^{m}$
instead of $h^{2m}$. 

For general $g$, we can only guarantee that 
\[
c\left(\frac{\prod_{(i,j)\in\gamma}\ez X\left[\pt{ij}\left(\X i,\X j\right)\right]}{\ez X\left[\pt X(\mathbf{X})\right]\prod_{k\in\beta}\ez X\left[\pt k\left(\X k\right)\right]}\right)=c\left(\frac{p'(\mathbf{X})}{p(\mathbf{X})}\right)+o(1),
\]
where $c(t_{1},t_{2})$ is a functional of the derivatives of $g$.
Applying this gives the final result in this case. However, we can
obtain higher order terms by making stronger assumptions on the functional
$g$ and its derivatives. Specifically, if $c(t_{1},t_{2})$ includes
functionals of the form of $t_{1}^{\alpha}t_{2}^{\beta}$ with $\alpha,\beta<0$,
then we can apply the generalized binomial theorem to use the higher
order expressions in (\ref{eq:varterms1}) and (\ref{eq:varterms2}).
This completes the proof.

\section{Proof of Variance Results}

To bound the variance of the plug-in estmiator, we use the Efron-Stein
inequality~\cite{efron1981jackknife}: 
\begin{lem}
(Efron-Stein Inequality) Let $\mathbf{X}_{1},\dots,\mathbf{X}_{n},\mathbf{X}_{1}^{'},\dots,\mathbf{X}_{n}^{'}$
be independent random variables on the space $\mathcal{S}$. Then
if $f:\mathcal{S}\times\dots\times\mathcal{S}\rightarrow\mathbb{R}$,
we have that 
\[
\var\left[f(\mathbf{X}_{1},\dots,\mathbf{X}_{n})\right]\leq\frac{1}{2}\sum_{i=1}^{n}\bE\left[\left(f(\mathbf{X}_{1},\dots,\mathbf{X}_{n})-f(\mathbf{X}_{1},\dots,\mathbf{X}_{i}^{'},\dots,\mathbf{X}_{n})\right)^{2}\right].
\]
\end{lem}
The Efron-Stein inequality bounds the variance by the sum of the expected
squared difference between the plug-in estimator with the original
samples and the plug-in estimator where one of the samples is replaced
with another iid sample.

In our case, consider the sample sets $\left\{ \mathbf{X}_{1},\dots,\mathbf{X}_{n}\right\} $
and $\left\{ \mathbf{X}_{1}^{'},\mathbf{X}_{2}\dots,\mathbf{X}_{n}\right\} $
and denote the respective plug-in estimators as $\gt$ and $\gt^{'}$.
Using the triangle inequality, we have 
\begin{align}
\left|\gt-\gt^{'}\right| & \leq\frac{1}{N}\left|g\left(\frac{\pt X^{'}(\mathbf{X}_{1})}{\pt X(\mathbf{X}_{1})}\right)-g\left(\frac{\pt X^{'}(\mathbf{X}_{1}^{'})}{\pt X(\mathbf{X}_{1}^{'})}\right)\right|+\frac{1}{N}\sum_{j=2}^{N}\left|g\left(\frac{\pt X^{'}(\mathbf{X}_{j})}{\pt X(\mathbf{X}_{j})}\right)-g\left(\frac{\left(\pt X^{'}(\mathbf{X}_{j})\right)^{'}}{\left(\pt X(\mathbf{X}_{j})\right)^{'}}\right)\right|,\label{eq:gJensen}
\end{align}
where $\left(\pt X^{'}(\mathbf{X}_{j})\right)^{'}$ and $\left(\pt X(\mathbf{X}_{j})\right)^{'}$
are the respective KDEs with $\mathbf{X}_{1}$ replaced with $\mathbf{X}_{1}^{'}$.
Then since $g$ is Lipschitz continuous with constant $C_{g}$, we
can write 
\begin{align*}
\left|g\left(\frac{\pt X^{'}(\mathbf{X}_{1})}{\pt X(\mathbf{X}_{1})}\right)-g\left(\frac{\pt X^{'}(\mathbf{X}_{1}^{'})}{\pt X(\mathbf{X}_{1}^{'})}\right)\right| & \leq C_{g}\left|\prod_{(i,j)\in\gamma}\pt{ij}\left(\X i_{1},\X j_{1}\right)-\prod_{(i,j)\in\gamma}\pt{ij}\left(\Xp i_{1},\Xp j_{1}\right)\right|\\
 & +C_{g}\left|\pt X(\mathbf{X}_{1})\prod_{k\in\beta}\pt k\left(\X k_{1}\right)-\pt X(\mathbf{X}_{1}^{'})\prod_{k\in\beta}\pt k\left(\Xp k_{1}\right)\right|.
\end{align*}
To bound the expected squared value of these terms, we split the product
of KDEs into separate cases. For example, if we consider the case
where the KDEs are all evaluated at the same point which occurs $M$
times, we obtain 
\[
\frac{M}{\left(Mh^{2}\right)^{2|\gamma|}}\sum_{m=2}^{N}\bE\left[\left(\prod_{(i,j)\in\gamma}K_{i}\left(\frac{\X i_{1}-\X i_{m}}{h}\right)K_{j}\left(\frac{\X j_{1}-\X j_{m}}{h}\right)-\prod_{(i,j)\in\gamma}K_{i}\left(\frac{\Xp i_{1}-\X i_{m}}{h}\right)K_{j}\left(\frac{\Xp j_{1}-\X j_{m}}{h}\right)\right)^{2}\right]
\]
\[
\leq\frac{1}{M^{2}}\prod_{(i,j)\in\gamma}||K_{i}K_{j}||_{\infty}^{2}.
\]
By considering the other $|\gamma|-1$ cases where the KDEs are evaluated
at different points (e.g. 2 KDEs evaluated at the same point while
all others are evaluated at different points, etc.), applying Jensen's
inequality gives 
\[
\bE\left[\left|\prod_{(i,j)\in\gamma}\pt{ij}\left(\X i_{1},\X j_{1}\right)-\prod_{(i,j)\in\gamma}\pt{ij}\left(\Xp i_{1},\Xp j_{1}\right)\right|^{2}\right]\leq C_{1}\prod_{(i,j)\in\gamma}||K_{i}K_{j}||_{\infty}^{2},
\]
where $C_{1}<\infty$ is some constant that is $O(1)$. Similarly,
we obtain 
\[
\bE\left[\left|\pt X(\mathbf{X}_{1})\prod_{k\in\beta}\pt k\left(\X k_{1}\right)-\pt X(\mathbf{X}_{1}^{'})\prod_{k\in\beta}\pt k\left(\Xp k_{1}\right)\right|^{2}\right]\leq C_{2}||K||_{\infty}^{2}\prod_{k\in\beta}||K_{k}||_{\infty}^{2}.
\]
Combining these results gives 
\begin{equation}
\bE\left[\left|g\left(\frac{\pt X^{'}(\mathbf{X}_{1})}{\pt X(\mathbf{X}_{1})}\right)-g\left(\frac{\pt X^{'}(\mathbf{X}_{1}^{'})}{\pt X(\mathbf{X}_{1}^{'})}\right)\right|^{2}\right]\leq C_{3},\label{eq:efron_term1}
\end{equation}
where $C_{3}=O(1)$.

As before, the Lipschitz condition can be applied to the second term
in (\ref{eq:gJensen}) to obtain 
\begin{align*}
\left|g\left(\frac{\pt X^{'}(\mathbf{X}_{m})}{\pt X(\mathbf{X}_{m})}\right)-g\left(\frac{\left(\pt X^{'}(\mathbf{X}_{m})\right)^{'}}{\left(\pt X(\mathbf{X}_{m})\right)^{'}}\right)\right| & \leq C_{g}\left|\prod_{(i,j)\in\gamma}\pt{ij}\left(\X i_{m},\X j_{m}\right)-\prod_{(i,j)\in\gamma}\pt{ij}^{'}\left(\X i_{m},\X j_{m}\right)\right|\\
 & +C_{g}\left|\pt X(\mathbf{X}_{m})\prod_{k\in\beta}\pt k\left(\X k_{m}\right)-\pt X^{'}(\mathbf{X}_{m})\prod_{k\in\beta}\pt k^{'}\left(\X k_{m}\right)\right|.
\end{align*}
For the first term, we again consider the $|\gamma|$ cases where
the KDEs are evaluated at different points. As a concrete example,
consider the example given in (\ref{eq:example}). Then we can write
by the triangle inequality
\[
\left|\prod_{(i,j)\in\gamma}\pt{ij}\left(\X i_{m},\X j_{m}\right)-\prod_{(i,j)\in\gamma}\pt{ij}^{'}\left(\X i_{m},\X j_{m}\right)\right|
\]
\begin{align*}
 & \leq\frac{1}{M^{2}h^{4}}\left|K_{1}\left(\frac{\X 1_{m}-\X 1_{1}}{h}\right)K_{2}^{2}\left(\frac{\X 2_{m}-\X 2_{1}}{h}\right)K_{3}\left(\frac{\X 3_{m}-\X 3_{1}}{h}\right)-K_{1}\left(\frac{\X 1_{m}-\Xp 1_{1}}{h}\right)K_{2}^{2}\left(\frac{\X 2_{m}-\Xp 2_{1}}{h}\right)K_{3}\left(\frac{\X 3_{m}-\Xp 3_{1}}{h}\right)\right|\\
 & +\frac{1}{M^{2}h^{4}}\left(\sum_{\substack{n=2\\
n\neq m
}
}^{N}K_{1}\left(\frac{\X 1_{m}-\X 1_{n}}{h}\right)K_{2}\left(\frac{\X 2_{m}-\X 2_{n}}{h}\right)\left|K_{2}\left(\frac{\X 2_{m}-\X 2_{1}}{h}\right)K_{3}\left(\frac{\X 3_{m}-\X 3_{1}}{h}\right)\right.\right.\\
 & \left.-K_{2}\left(\frac{\X 2_{m}-\Xp 2_{1}}{h}\right)K_{3}\left(\frac{\X 3_{m}-\Xp 3_{1}}{h}\right)\right|+\left|K_{1}\left(\frac{\X 1_{m}-\X 1_{1}}{h}\right)K_{2}\left(\frac{\X 2_{m}-\X 2_{1}}{h}\right)-K_{1}\left(\frac{\X 1_{m}-\Xp 1_{1}}{h}\right)K_{2}\left(\frac{\X 2_{m}-\Xp 2_{1}}{h}\right)\right|\\
 & \left.\times\sum_{\substack{n=2\\
n\neq m
}
}^{N}K_{2}\left(\frac{\X 2_{m}-\X 2_{n}}{h}\right)K_{3}\left(\frac{\X 3_{m}-\X 3_{n}}{h}\right)\right).
\end{align*}
\begin{equation}
\implies\bE\left[\left|\prod_{(i,j)\in\gamma}\pt{ij}\left(\X i_{m},\X j_{m}\right)-\prod_{(i,j)\in\gamma}\pt{ij}^{'}\left(\X i_{m},\X j_{m}\right)\right|^{2}\right]\leq\frac{4+6(M-2)^{2}}{M^{4}}||K_{1}K_{2}^{2}K_{3}||_{\infty}^{2}.\label{eq:kde_diff}
\end{equation}
For more general $\gamma$, it can be shown that the LHS of (\ref{eq:kde_diff})
is $O\left(\frac{1}{M^{2}}\right)$. Similarly, we can check that
\[
\bE\left[\left|\pt X(\mathbf{X}_{m})\prod_{k\in\beta}\pt k\left(\X k_{m}\right)-\pt X^{'}(\mathbf{X}_{m})\prod_{k\in\beta}\pt k^{'}\left(\X k_{m}\right)\right|^{2}\right]=O\left(\frac{1}{M^{2}}\right).
\]

Applying the Cauchy-Schwarz inequality with these results then gives
\begin{equation}
\bE\left[\left(\sum_{j=2}^{N}\left|g\left(\frac{\pt X^{'}(\mathbf{X}_{j})}{\pt X(\mathbf{X}_{j})}\right)-g\left(\frac{\left(\pt X^{'}(\mathbf{X}_{j})\right)^{'}}{\left(\pt X(\mathbf{X}_{j})\right)^{'}}\right)\right|\right)^{2}\right]=O(1).\label{eq:efron_term2}
\end{equation}
Combining (\ref{eq:efron_term1}) and (\ref{eq:efron_term2}) with
(\ref{eq:gJensen}) gives 
\[
\bE\left[\left|\gt-\gt^{'}\right|^{2}\right]=O\left(\frac{1}{N^{2}}\right).
\]
Applying the Efron-Stein inequality then gives 
\[
\var\left[\gt\right]=O\left(\frac{1}{N}\right).
\]

\end{document}